\documentclass[12pt,preprint]{aastex}

\font\syvec=cmbsy10
\newcommand\bnabla{\hbox{{\syvec\char114}}}       
\font\tenbsy=cmbsy10
\newcommand\bdot{{\hbox{\tenbsy \char'1}}}
\newcommand\ie{{\it i.e.}}
\newcommand\eg{{\it e.g.}}

\newcommand\Msun{M_\odot}
\newcommand\Lsun{L_\odot}

\newcommand\etal{{\it et al.~\/}}
\newcommand\lta{\lower.5ex\hbox{\ltsima}}
\newcommand\gta{\lower.5ex\hbox{\gtsima}}
\newcommand\ltsima{$\; \buildrel < \over \sim \;$}
\newcommand\lsim{\lower.5ex\hbox{\ltsima}}
\newcommand\gtsima{$\; \buildrel > \over \sim \;$}
\newcommand\gsim{\lower.5ex\hbox{\gtsima}}

\shorttitle{Star Formation in Space and Time: The Orion Nebula Cluster}
\shortauthors{Eric M. Huff and Steven W. Stahler}

\begin{document}
\title{Star Formation in Space and Time: \\
The Orion Nebula Cluster}
\author{E. M. Huff and Steven W. Stahler}
\affil{Astronomy Department, University of California, Berkeley, CA 94270}

\begin{abstract}
We examine the pattern of star birth in the Orion Nebula Cluster (ONC), 
with the goal of discerning the cluster's formation mechanism. Outside of the Trapezium,
the distribution of stellar masses is remarkably uniform, and is not accurately
described by the field-star initial mass function. The deconvolved, 
three-dimensional density of cluster members peaks at the Trapezium stars,
which are truly anomalous in mass. Using theoretical pre-main-sequence tracks,
we confirm the earlier finding that star formation has accelerated over the 
past $10^7$~yr. We further show that the rate of acceleration has been the same
for all masses. Thus, there is no correlation between stellar age and mass,
contrary to previous claims. Finally, the acceleration has been spatially
uniform throughout the cluster.

Our reconstruction of the parent molecular cloud spawning the cluster shows
that it had a mass of $6700\,\,\Msun$ prior to its destruction by the Trapezium. 
If the cloud was supported against self-gravity by mildly dissipative 
turbulence, then it contracted in a quasi-static, but accelerating manner. We
demonstrate this contraction theoretically through a simple energy argument. 
The mean turbulent speed increased to its recent value, which 
is reflected in the present-day stellar velocity dispersion. 

The current ONC will be gravitationally unbound once cloud destruction is
complete, and is destined to become a dispersing OB association. We hypothesize
that similarly crowded groups seen at the centers of distant OB associations 
are also unbound, and do not give rise to the Galactic population of open 
clusters. Finally, accelerating star formation implies that most clumps
within giant molecular complexes should have relatively low formation activity.
Sensitive infrared surveys could confirm this hypothesis.  
\end{abstract}

\bigskip

\section{Introduction}
The origin of stellar groups is one of the abiding mysteries of astronomy. 
Infrared and radio observations have demonstrated that these aggregates are 
born within cold molecular clouds (Lada \& Lada~2003). We also know that young 
groups come in three varieties~-- T~associations, open clusters, and OB 
associations. Yet we have little notion as to what distinguishes the progenitor 
cloud for each type.

The statistics of stellar births on the Galactic scale indicates that most 
stars originate in OB associations (Roberts~1957; Miller \& Scalo~1978). These 
populous groups form inside clumps within giant molecular complexes  
(Blitz~1993). The parent clump must have attained a high density at the 
formation site of the massive O and B stars, since the youngest of these are 
almost always found in crowded stellar fields (e.g., Stahler \etal~2000). But 
how did the cloud attain such a high density? More generally, what forces 
drove its evolution as it was producing the association?

By far the best-studied young OB association is the Orion Nebula Cluster. Over
a thousand members are seen only in the near-infrared (Ali \& Depoy 1995).
However, a comparable number are optically visible (\eg, Prosser \etal 1994), 
since much of the parent cloud is now dispersed. Foreground neutral gas is 
largely confined to a relatively thin layer (O'Dell \etal 1992), while gas 
ionized by $\theta^1$~Ori~C and its massive companions is being driven outward
(Gordon \& Churchwell 1970; O'Dell 1994). Behind this hemispherical blister of
stars and ionized gas lies a dense, massive wall of molecular gas, still 
creating new stars (see, \eg, Genzel \& Stutzki 1989).

In this paper, we examine anew both the present morphology of the cluster and
its past record of star formation. Our goal is to elucidate, as much as 
possible, the properties and dynamical history of the parent cloud body. This 
contribution is the latest in a continuing investigation of stellar groups and
their origin. Palla \& Stahler (1999; hereafter Paper~I) found that star 
formation in this region has been accelerating over time, a result seen 
generally, albeit at a reduced level, in other nearby groups (Palla \& Stahler
2000; Paper~II). In a study of pre-main-sequence binaries, Palla \& Stahler 
(2001) showed that the companion to BM~Ori is no older than $10^5$~yr. It thus
apears that the centrally located Trapezium stars were formed relatively 
recently. Here, we wish to provide a fuller account of the region's evolution 
in both space and time, and to draw implications for stellar group formation 
generally.

Our main results may be readily summarized. The Trapezium stars are truly
anomalous within the group, with a total mass equivalent to hundreds
of ordinary stars. Excluding the Trapezium, the distribution of stellar masses is nearly uniform 
throughout the remaining cluster. The fractional increase with time in the 
stellar population has also been the same at all locations. In other words, the 
acceleration in starbirth, which occurred over some $10^7$~yr, was a 
{\it global} phenonemon, and resulted in a remarkably homogeneous membership, 
except at the very center.

We hypothesize that this rise in star formation activity was stimulated by
large-scale contraction of the ONC parent cloud. The gas density must have
increased in a quasi-static, but accelerating manner. We explore a simple
model, in which the cloud was supported against self-gravity by turbulent 
pressure. The energy associated with this turbulent motion is continually 
dissipated by internal shocks. In our model, the mean turbulent speed 
{\it increases} with time because of gravitational compression. Energy 
dissipation from shocks also increases, which allows faster contraction. In 
summary, our model illustrates how accelerating contraction, along with a 
concurrent rise in star formation, is a natural and generic consequence of 
support from internal turbulence.   

Section 2 below presents our analysis of the present-day ONC stellar 
distribution. In Section 3, we utilize pre-main-sequence stellar ages to infer
the star formation history throughout the region. Section 4 offers our 
physical interpretation of this history in terms of the parent cloud and its 
evolution. Finally (\S 5), we discuss the implications of this study, both for
the ONC and for stellar group formation generally.

\section{Current Stellar Population}
\subsection{Density Variation}
Our empirical dataset comes from the important study by Hillenbrand (1997). 
Combining her own observations with the existing literature, Hillenbrand
obtained both optical spectra and $V$- and $I$-band fluxes for 934 stars within 
2.5~pc of the Trapezium. A comparable number of stars are detectable only at
near-infrared wavelengths (Ali \& Depoy 1995). The optical spectra yielded
spectral types, which in turn gave effective temperatures. Observed $V$-$I$ 
colors were then employed to find individual extinctions. A bolometric 
correction to the $I$-magnitudes, together with the extinction, then gave 
stellar luminosities. In this manner, the stars were placed in the theoretical
HR diagram, \ie, the $L_\ast$-$T_{\rm eff}$ plane.

Figure~1 shows the positions of a large subsample, 705 stars, from the full,
optically visible set. These are the objects that, according to Hillenbrand, 
have a membership probability exceeding 67~percent. These probabilities were 
assessed from earlier proper-motion studies, principally that of Jones \& 
Walker (1988). As is well known, the stars are crowded toward 
$\theta^1$~Ori~C, whose own position is central in the plot. The conspicuous 
gap in stars just to the east of the Trapezium coincides with the Dark Bay,
familiar from optical photographs of the region (\eg, Pogge et al. 1992).

\begin{figure}[!h]
\epsscale{0.7}
\plotone{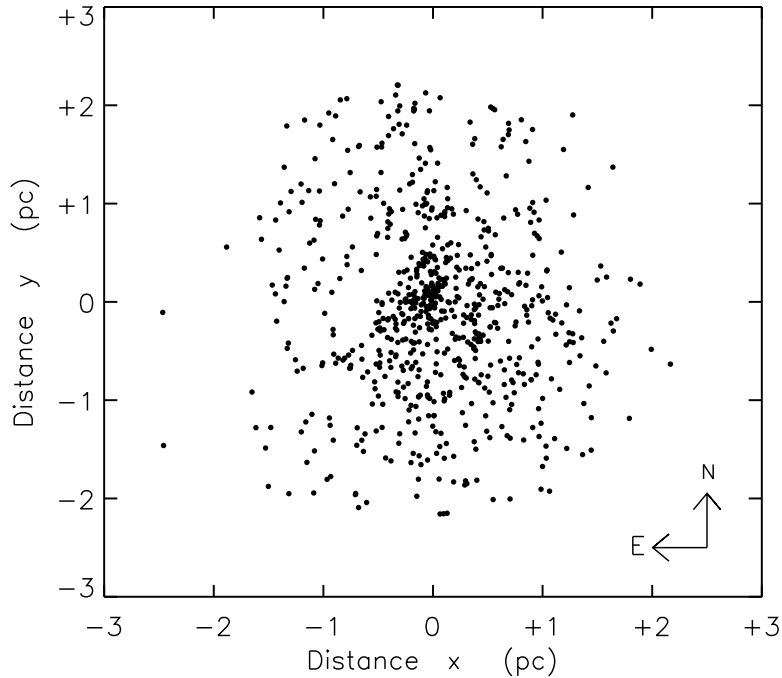}
\caption{Positions in the sky of the 705 ONC stars with
67~percent membership probability (from Hillenbrand 1997). The $x$- and
$y$-axis are oriented in the directions of right ascension and declination,
respectively, and a distance of 480~pc is assumed.}
\end{figure}

What is the true, three-dimensional distribution of these stars? It has long
been accepted that the visible population is contained within an ionized blister
of gas, located in front of OMC-1, itself a portion of the extended Orion~A
cloud (Zuckerman 1973). Wen \& O'Dell (1995) used the distribution in 
H$\alpha$ emission to map out the ionized front surface of OMC-1. They found 
that the main ionizing source, $\theta^1$~Ori~C, is situated only 0.2~pc from 
this rear wall. (See their Figure 3.) This distance is small compared to the 
full cluster diameter of 5~pc. Hence we may assume, with little loss of 
precision, that the ONC morphology is hemispherical.\footnote{The full 
cluster morphology may be spherical. If so, the many near-infrared sources 
imaged by Ali \& Depoy (1995) and by Hillenband \& Carpenter (2000) plausibly
represent the rear half of the sphere, \ie, that fraction partially embedded in 
the OMC-1 wall.}   

Working within this model, we first azimuthally average the projected 
stellar distribution, again choosing $\theta^1$~Ori~C as our origin. In Figure~2
we show this empirical distribution as a function of the radius $r$. The
quantity displayed is the number of stars enclosed at each $r$-value. This
cumulative number is then converted to a fraction of the total stellar 
population, and denoted as $f_\ast (r)$. 

\begin{figure}[!h]
\epsscale{0.7}
\plotone{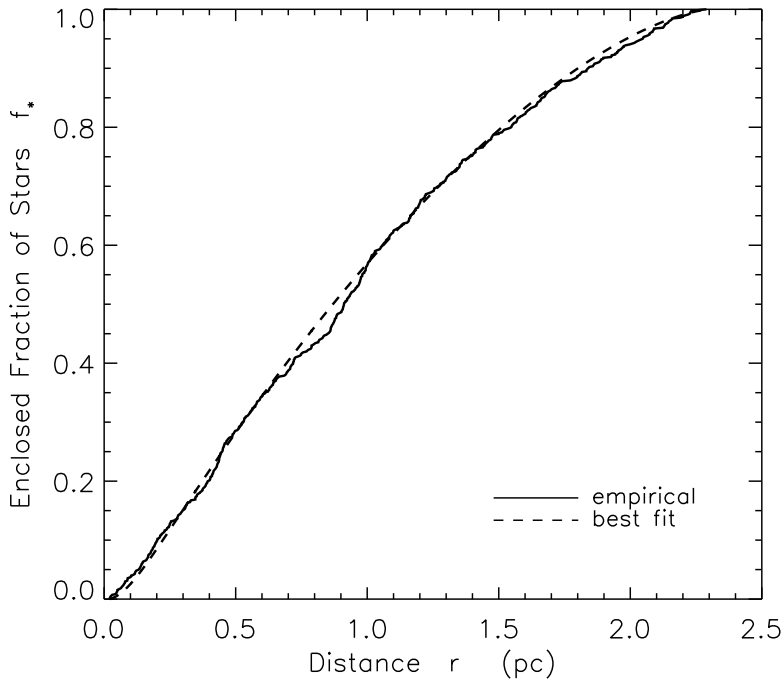}
\caption{The projected number density of stars. The fraction of
the total stellar population of 705 stars is shown as a function of projected 
radius. The dashed curve is a smooth fit to the data.}
\end{figure}

We fit these data by adopting a simple form for the three-dimensional stellar
number density throughout the hemisphere:
$$ n_\ast (r)\,=\, n_c\,\left[1\,-\,{r\over r_1}\,+\,
             \left({r\over r_2}\right)^{\!\!2}\right]^{\!-1} \,\,.\eqno(1) $$
For each choice of the scale lengths $r_1$ and $r_2$, we project this density
onto the plane of the sky and compute the cumulative number fraction 
$f_\ast (r)$. We then vary $r_1$ and $r_2$ until our model $f_\ast (r)$ best 
matches the empirical curve. Note the negative sign preceding the term 
containing $r_1$; this choice is necessary to match the relatively rapid rise 
in the observed $f_\ast (r)$ away from the origin. Note also that the central 
density $n_c$ is unconstrained by this matching procedure.

The smooth curve in Figure~2 shows the best-fit $f_\ast (r)$, obtained by
selecting \hbox{$r_1\,=\,0.25\ {\rm pc}$} and \hbox{$r_2\,=\,0.14\ {\rm pc}$}.
To estimate the central density, we integrate equation (1) over the entire
hemisphere and equate the total number of stars to the observed 705. We thus
find \hbox{$n_c\,=\,2300\ {\rm pc}^{-3}$}. This result is close to that given
by Herbig \& Terndrup (1986), who surveyed the inner 0.5~pc in radius. Both
numbers are dwarfed by the density of $5\times 10^4$~pc$^{-3}$ found in the 
high-resolution, near-infrared study of McCaughrean \& Stauffer (1994). Their
high value was obtained by counting stars within 0.05~pc of $\theta^1$~Ori~C, 
a distance comparable to that of the other Trapezium members.   

It is interesting that the stellar number density falls off as $r^{-2}$ for
\hbox{$r\,\gg\,r_2$}. Of course, we forced this result mathematically through
our adopted functional form of $n_\ast (r)$. We were guided in this choice by
previous authors, who noted that the {\it projected} areal density falls 
approximately as $r^{-1}$ (see, \eg, Scally \& Clarke 2001). How robust is this
result? We have generalized equation (1) to include an exponent $2+\epsilon$ 
in the term involving $r_2$. We then varied $\epsilon$ along with $r_1$ and 
$r_2$ in our matching procedure. An acceptable fit, in the least-squares sense,
could only be obtained for $\epsilon$ ranging from -0.1 to +0.4.

If this $r^{-2}$ falloff continued to very large radii, the cumulative, 
projected starcount would increase as $r$. According to Figure 2, the rise is 
actually slower, both observationally and in the theoretical fit. We are seeing
the effect of a finite boundary, corresponding to 2.5~pc in radius. Outside 
this boundary, no stars are counted.\footnote{A singular isothermal 
sphere truncated at radius $R$ has a projected surface density which varies as
$r^{-1}$ times the correction factor 
\hbox{$\,{\rm arctan}\sqrt{(R/r)^2\,-\,1}$.}} Of course, the stellar population
does not truly vanish, but there is a steeper falloff in this vicinity, as 
the central group blends into the larger Orion~Ic association (Herbig \& 
Terndrup 1986).

Figure~3 displays the reconstructed, three-dimensional density $n_\ast (r)$.
One notable feature here is the central dip. However, the results of 
McCaughrean \& Stauffer (1994) indicate that this falloff, which occurs for
\hbox{$r\,\lsim\,0.05\ {\rm pc}$}, is not real. The true stellar density climbs
steeply toward the center. At visible wavelengths, this interior population is
lost in the brilliant glare of the Trapezium. It is at least partially recovered
in high-resolution, near-infrared imaging. (See also Hillenbrand \& Carpenter
2000.) The dashed curve in Figure~3 is a pure $r^{-2}$ profile. Extended inward,
its rise is suggestive of that found empirically within 0.05~pc.

\begin{figure}[!h]
\epsscale{0.7}
\plotone{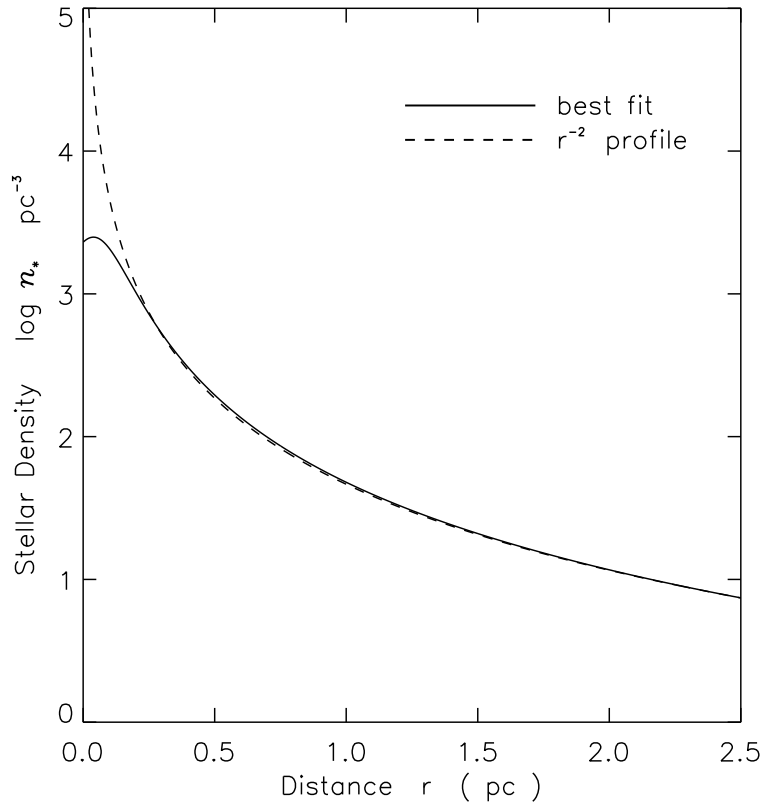}
\caption{Reconstructed, three-dimensional density of stars,
compared to a $r^{-2}$ profile ({\it dashed curve}). The inward falloff in the 
reconstructed density is not real, but reflects incompleteness of the optical
sample in the crowded central region.}
\end{figure}

\subsection{Mass Distribution}
We turn next to the spatial distribution of stellar masses in the present-day
cluster. To obtain a full sampling of masses among the optical population,
we need to consider the sensitivity limit of the original survey. Here we are
concerned with spectroscopic observations, which are needed to place stars in 
the HR diagram. 

According to Hillenbrand (1997), her observations are complete to 
\hbox{$I\,=\, 17$}. (See upper panel of her Figure 6.) A typical ONC star is
of spectral type M2, for which \hbox{$V-I\,=\,2.4$}. The bolometric correction 
needed to extrapolate from the $V$-band to $M_{\rm bol}$ is $-1.8~{\rm mag}$ 
(Hillenbrand 1997; Appendix C). For the representative extinction 
\hbox{$A_V\,=\,2$}, and the distance modulus of 8.4~mag, we find that the 
limiting $I$-magnitude corresponds to a luminosity of $0.1\ \Lsun$.   

We may now address the completeness of stellar masses. Suppose we limit our
investigation to stars with ages less than $1\times 10^7$~yr. Then the
lowest stellar mass for which we have complete data is that whose 
pre-main-sequence track descends to 0.1~$\Lsun$ at $t = 1\times 10^7$~yr. 
Inspection of Figure~1 from Palla \& Stahler (1999) shows that this critical
mass is $0.4\ \Msun$. When discussing stellar masses and ages, we will
henceforth limit our attention to the restricted sample of 244 stars with 
\hbox{$M_\ast \ge 0.4\ \Msun$}.    

It is important to note that the failure to account for survey completeness
may lead to puzzling and erroneous results. For example, Hillenbrand (1997) 
observed that lower-mass stars in both Orion and other regions are younger, on
average, than their high-mass counterparts. This age discrepancy was noted 
again by Hartmann (2004), and led him to reject the pre-main-sequence tracks 
themselves as reliable age indicators. However, the lower-mass population is 
incompletely sampled in all surveys. The missing objects are precisely those 
of lower luminosity, and hence greater ages. Within our more carefully defined 
sample, we find no age-mass correlation (see \S 3 below).

We now wish to explore both the distribution 
of masses and how this distribution varies spatially in the cluster. A simple
but instructive exercise is to divide the cluster into two groups of 
equal population. The inner group lies within a circle of radius 0.88~pc, 
centered on $\theta^1$~Ori~C. The outer group is in the surrounding annulus.
We rank stars in each group by mass, and calculate the fraction of the group
at or below every mass-value. To construct the figure, we find, for each mass
in one group, that mass in the other group which represents the same
cumulative fraction. (Such a {\it quantile-quantile plot} is a common
statistical tool for comparing populations; see Evans et al. 2000).

Figure~4 shows the result. The vast majority of stars match up, \ie, the two 
populations are essentially identical. (The step-like character of the 
empirical curve is an artifact of the discrete spectral typing.) Above 
\hbox{$M_\ast \approx 3 \ \Msun$}, however, the two populations differ 
radically. The inner group contains objects which are more massive than any 
found outside. Naturally, the Trapezium stars are part of this exceptional
subgroup. Figure~4, which displays masses on a logarithmic scale, shows
vividly how the Trapezium is truly anomalous. If we ignore its central region, 
the cluster appears to be homogeneous in terms of membership.

\begin{figure}[!h]
\epsscale{0.7}
\plotone{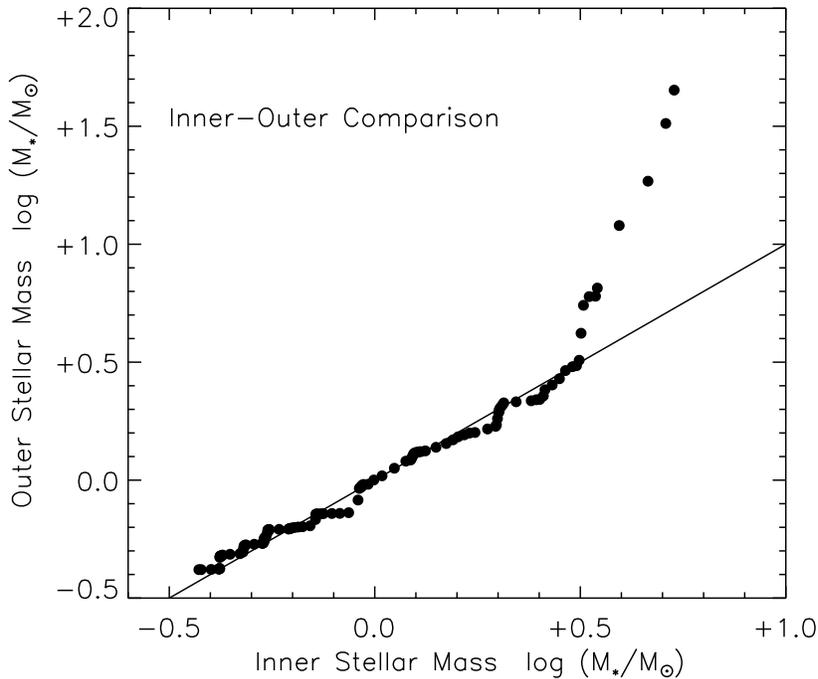}
\caption{Comparison of stellar masses in the inner and outer
halves of the cluster. Note the anomalously massive stars within the inner
region.}
\end{figure}

Another way to demonstrate this point is presented in Figure~5. The dashed 
curve in both panels represents the cumulative mass fraction, \ie, the
fraction of the total cluster mass contained in each radius $r$. The solid
curve is the cumulative number fraction. If we include the Trapezium 
({\it left panel}), the mass fraction immediately rises above the number
fraction. If we omit the Trapezium ({\it right panel}), the mass and number
fractions are nearly identical from the center outward. In other words, 
increasing the number of stars by a certain fraction gives the same fractional
increase in mass, at all radii. The distribution of individual stellar masses 
must therefore be similar throughout the cluster. Clearly, we are not seeing
mass segregation in the traditional sense.

\begin{figure}[!h]
\epsscale{0.9}
\plotone{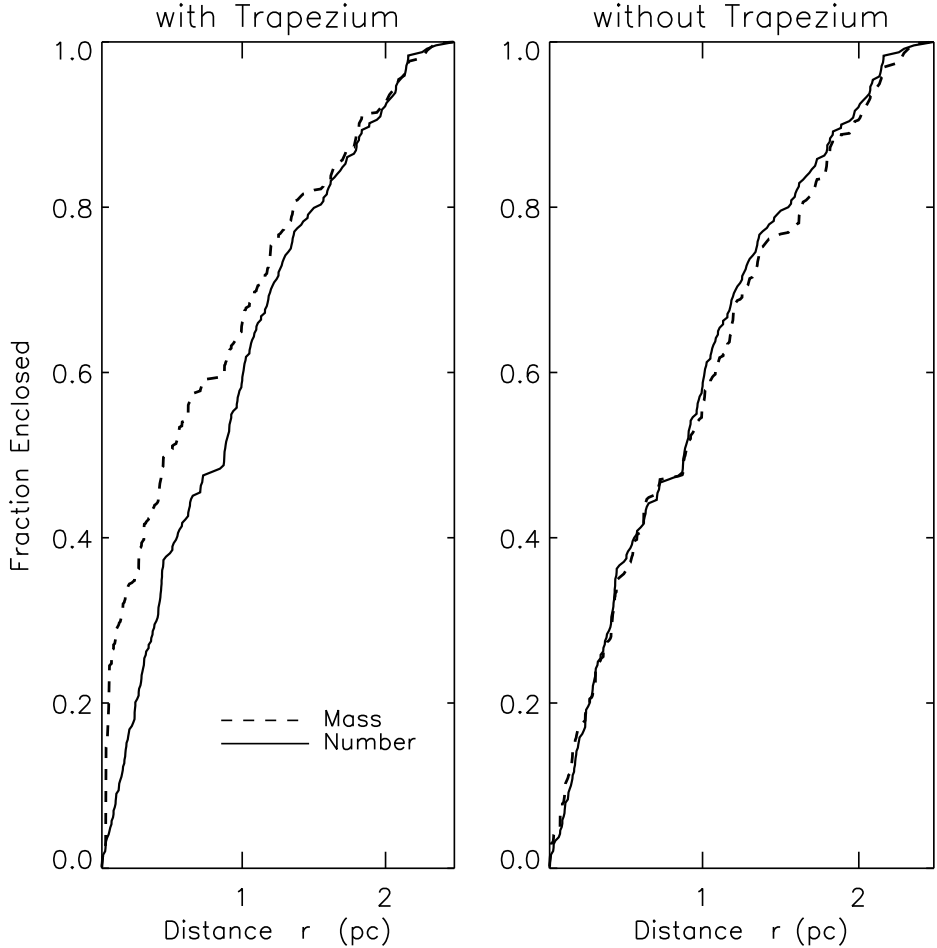}
\caption{ Comparison of the distributions of stellar mass and
number as a function of radius. {\it Left panel:} When the Trapezium is 
included, the mass curve immediately climbs higher than the stellar number.
{\it Right panel:} Without the Trapezium, the two distributions are nearly
identical, indicating uniformity of the stellar mass spectrum.}
\end{figure}

It also follows that the local mass distribution is essentially the same as the
global one. Paper~I (\S 4) asserted that the ONC stars roughly follow the
field star initial mass function. The top two curves of Figure~6 show that the
two distributions actually differ in a significant way. The solid curve 
displays the cumulative mass function ${\cal N}_\ast (M_\ast)$, \ie, the number
of stars with masses up to the value $M_\ast$. For comparison, the dashed curve
shows the same quantity calculated from the field star initial mass function 
(Scalo 1998). (This and the other dashed curves are normalized to the total
population in each age range). It is apparent that the ONC is deficient in members with 
\hbox{$0.8\ \Msun \lsim M_\ast \lsim 3.0\ \Msun$}. A similar discrepancy 
appears in the analysis of Hillenbrand (1997), although she used different 
pre-main-sequence tracks and a different field-star initial mass function. 
(See the upper panel of her Figure 15.)  The remaining curves of Figure~6 
concern the temporal development of the cluster, and it is this subject that 
we now consider.

\begin{figure}[!h]
\epsscale{0.7}
\plotone{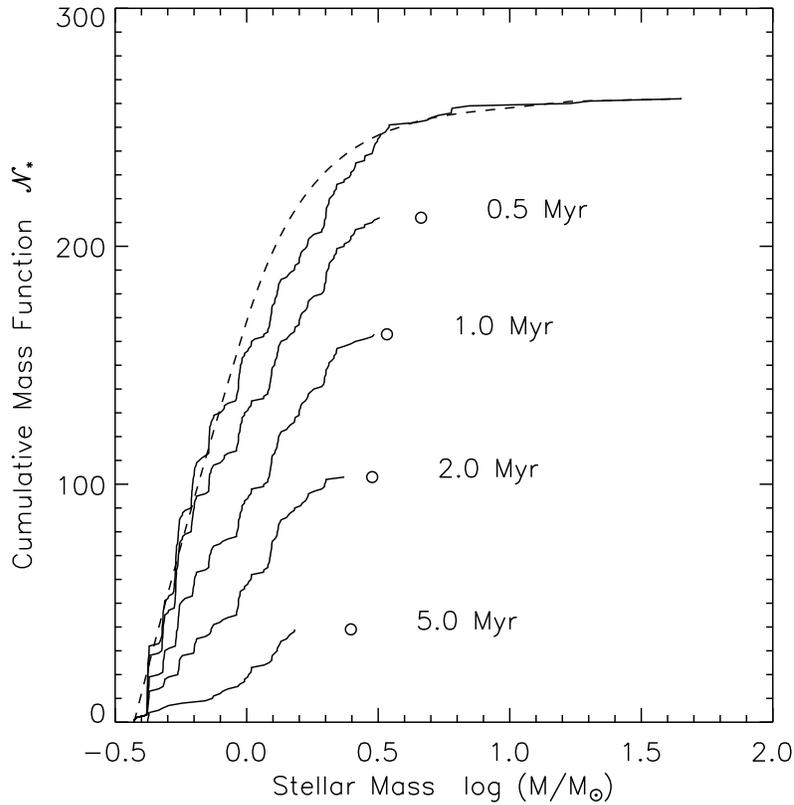}
\caption{The cumulative mass function for stars of different
ages. The bottom curve represents only stars older than 5~Myr; the next
curve is for stars older than 2~Myr, etc. The top curve shows this function 
for the entire sample of 244 stars. Each open circle marks the stellar mass 
whose pre-main-sequence lifetime is the value indicated. The dashed curve was 
constructed from the field-star initial mass function of Scalo (2002). }
\end{figure}

\section{Star Formation History}
The distribution of pre-main-sequence ages provides the essential record of
star formation in any group. Palla \& Stahler (1999) used this technique to 
show that production of stars within the ONC began some $10^7$~yr in the past 
and has been accelerating to the present. We now go one step further, and 
combine stellar ages and {\it locations}. In other words, we assess the 
evolution of the region both temporally and spatially.

The four panels of Figure~7 display maps of the ONC at various epochs. Within
each time interval (given in the caption), the panel shows all stars from the 
sample of 244 that were born during that interval. The small symbols denote 
low-mass objects \hbox{($M_\ast \, < \, 2\ \Msun$)}, while the larger ones 
represent those of intermediate mass \hbox{($2\ \Msun < M_\ast < 8\ \Msun$)}. 
We have omitted the three innermost Trapezium stars, which would not be 
resolved in these plots. None of these three have reliable ages, although they
were probably formed within the most recent epoch (Palla \& Stahler 2001). Note 
that the lowest-mass Trapezium star, BM~Ori, has 7~$\Msun$, and is thus 
included in the final map.

\begin{figure}[!h]
\epsscale{0.9}
\plotone{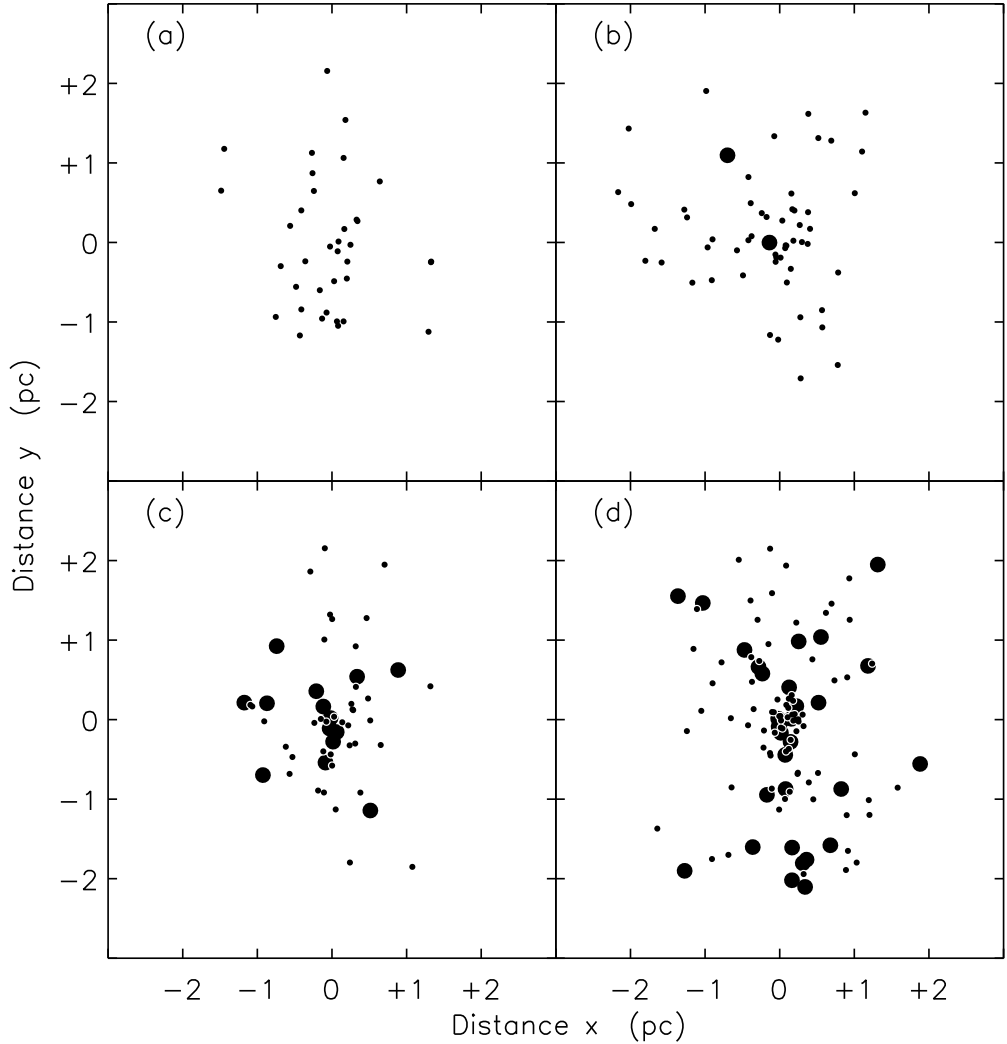}
\caption{Positions of ONC stars (from the sample of 244) that
were born (a) more than 5~Myr ago, (b) from 2 to 5~My ago, (c) from 1 to 2~Myr 
ago, and (d) more recently than 1~Myr. The smaller filled circles denote stars 
with \hbox{$M_\ast\,<\,2\,\Msun$}, while the larger ones are for \hbox{$2\ \Msun < M_\ast < 8\ \Msun$}}
\end{figure}

It is apparent that the earliest star formation, which occurred some $10^7$~yr
in the past, was diffuse spatially, and did not exhibit crowding toward the
future site of the Trapezium. Intermediate-mass stars also began forming early,
and were also spread throughout the region. As expected, the entire population
increased dramatically within the last few Myr. No star formation is occurring
now within the volume occupied by the visible stars, since molecular gas has
been driven off. If, however, the cluster extends into the wall of OMC-1, then
starbirth is undoubtedly continuing apace in that embedded region.

The ONC stars have transverse velocities of several km~s$^{-1}$ (Jones \&
Walker 1988). Over $10^7$~yr, they would have moved tens of parsecs, well out
of the area shown. How, then, can we identify their present positions with
their birth sites? The reason is that {\it the stars have not been moving on 
ballistic trajectories}. They have been subject to the gravitational pull from
other cluster members and, much more significantly, from the ambient
molecular gas. In other words, they were trapped in local potential wells until
recently, when the nearby gas was ionized and dispersed. If this dispersal
occurred $10^5$~yr ago, then the stars subsequently moved only a few tenths of 
a parsec.

We also arrive at this picture of the stellar motion by following the 
consequences of the alternative view. Suppose that all stars were born in 
the most crowded region, near the present-day Trapezium. Suppose further that 
they drifted away from their birth sites at constant speed. Then we would 
see today an age gradient, in the sense that the older stars would be
farther from the center. Figure~8 plots the mean stellar age at each projected
radius, along with the rms dispersion. The mean age does not increase outward,
but remains constant to within the statistical errors. Under the drift 
hypothesis, furthermore, proper motion vectors would tend to point outward for
more distant members. This effect is also not seen; the vectors are randomly 
oriented (Jones \& Walker 1988).

\begin{figure}[!h]
\epsscale{0.7}
\plotone{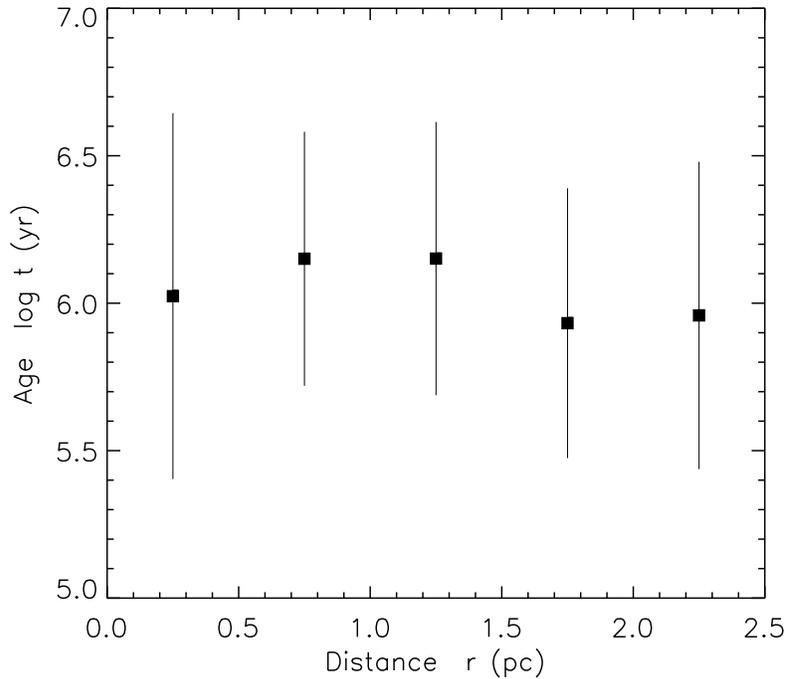}
\caption{ Pre-main-sequence stellar age as a function of 
projected radius. Shown is ${\rm log}\,t$, averaged within an annulus centered
on each displayed radius. The rms error is also shown at each position.}
\end{figure}

A plot of the mean age versus stellar mass is also of interest, 
especially in light of the previous claims for a correlation (Hillenbrand 1997;
Hartmann 2004). Figure 9 shows the result. When we limit ourselves to the sample 
of 244 stars, the mean pre-main-sequence age is essentially independent of mass, 
at least for \hbox{$M_\ast \lsim 2.5\ \Msun$}. For higher masses, the mean age does 
systematically fall, in apparent agreement with the earlier claims.
 
However, this falloff only reflects the diminishing pre-main-sequence lifetime
of more massive objects. The dashed curve displays this lifetime, \ie, the
contraction time from the birthline to the zero-age main sequence, as a function
of mass. Above $2.5~\Msun$, a significant fraction of the stars have already reached
the main sequence. However, all stars in the sample are assigned {\it pre}-main-sequence
ages, according to their positions on the HR diagram. The main-sequence stars
thus have ages which are too young. Since the relevant fraction increases with
stellar mass, there is a spurious correlation of age with mass.

\begin{figure}[!h]
\epsscale{0.7}
\plotone{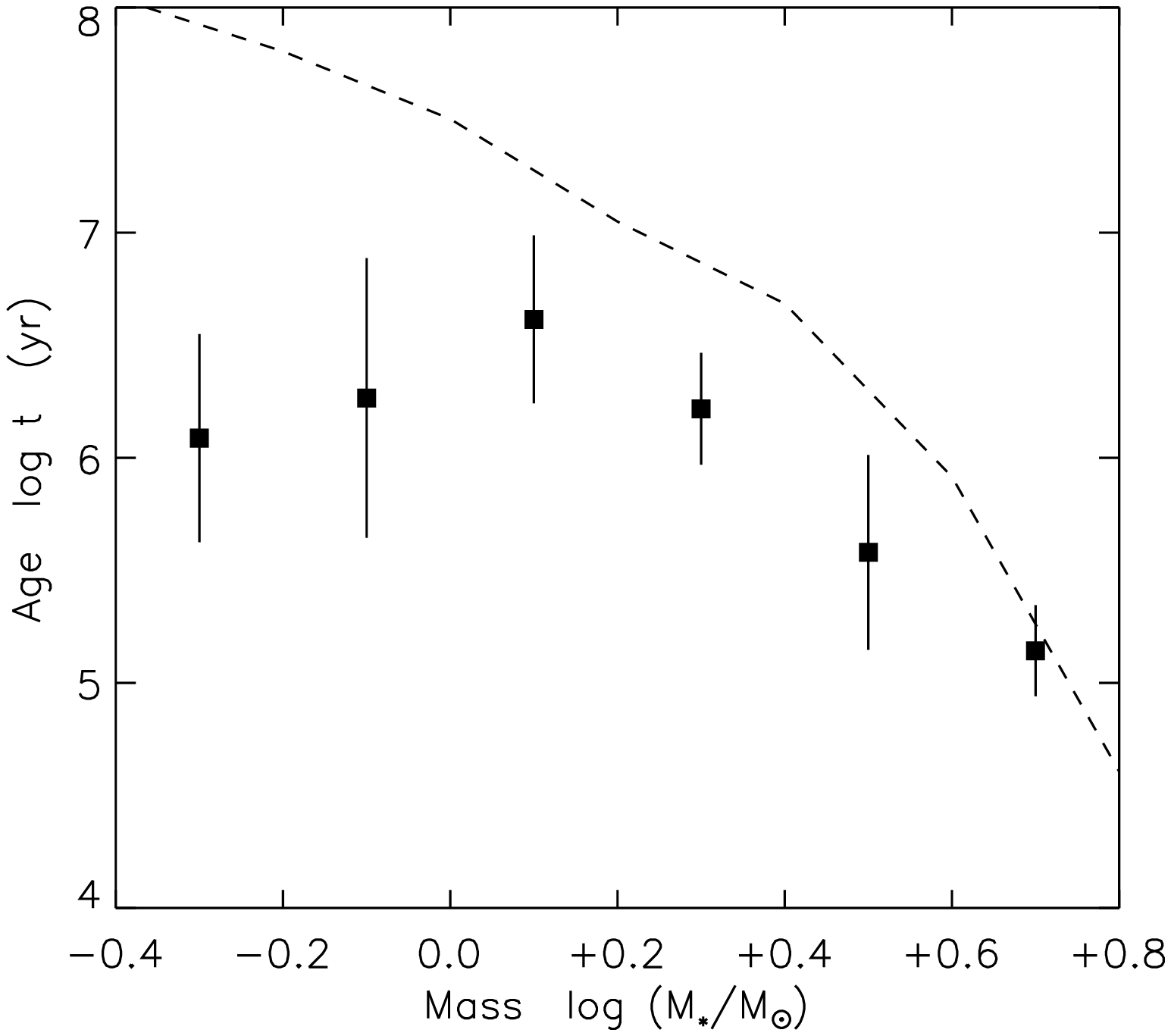}
\caption{Pre-main-sequence stellar age as a function of mass.
This plot was constructed in a manner analogous to Figure 8. The dashed curve
represents the total contraction time to the zero-age main sequence at each
mass value.}
\end{figure}

Returning to the global evolution, the sequential maps of Figure~7 demonstrate
that star formation is accelerating, but they do not tell us the spatial 
pattern of that acceleration. Is the increase of stellar births predominantly 
near the center? Has the acceleration process perhaps moved inward over time? 
Such details are important if we are to discern the physical processes at play
in group formation.

It is again instructive to divide the cluster into two groups of equal 
population, as we did when constructing Figure~4. This time, we compare the
star formation histories of the inner and outer portions of the cluster. The
result is shown in Figure~10. Here we display the fraction of stars in each
group that have at least the indicated ages. In other words, we are showing
that portion of each group formed at any time. The two curves are nearly
identical.

\begin{figure}[!h]
\epsscale{0.9}
\plotone{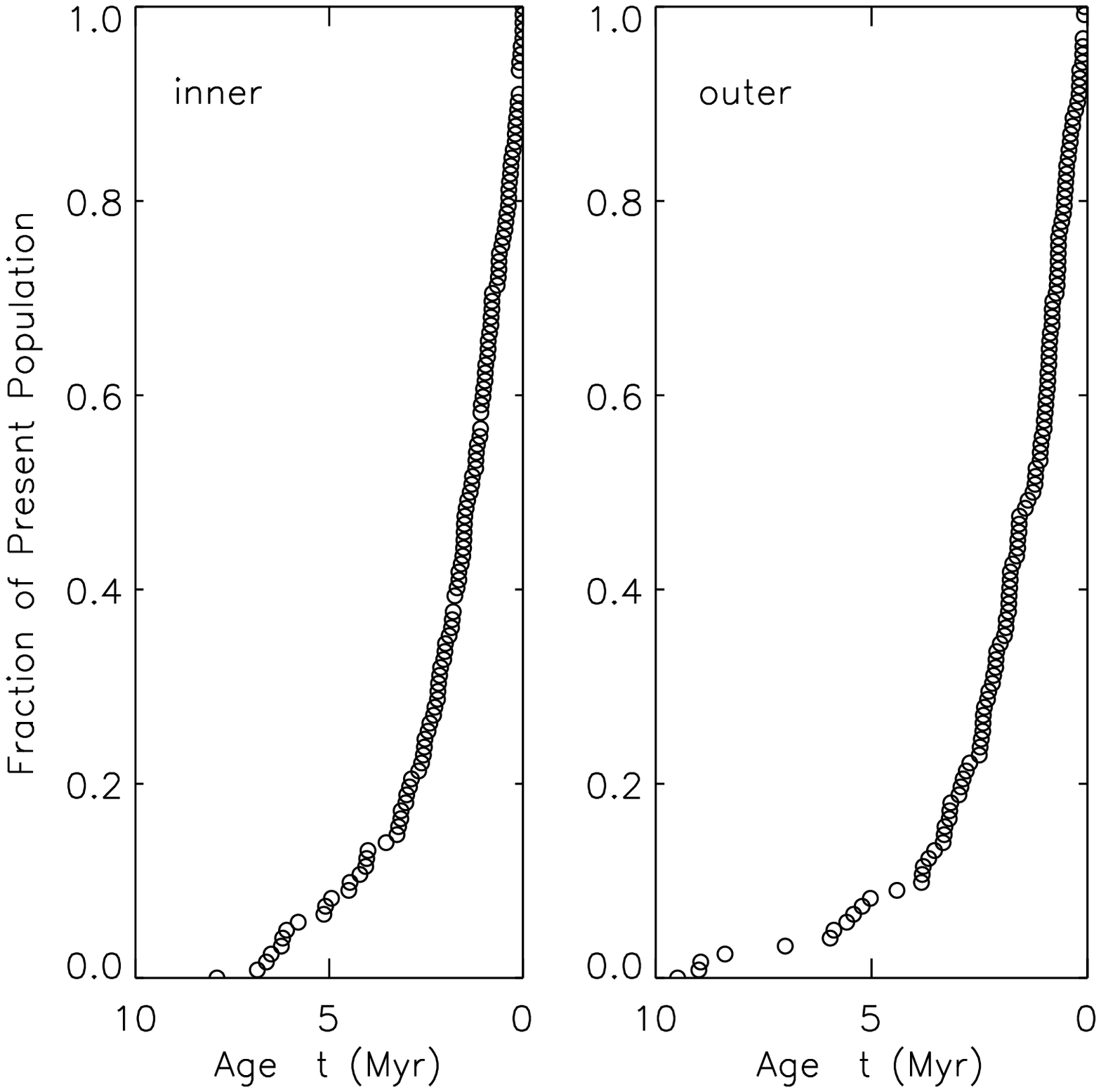}
\caption{Acceleration of starbirth in the inner and outer
regions of the cluster. Plotted in each panel is the fraction of the present
population that has the indicated age.}
\end{figure}

What we have found is that the acceleration did not occur locally. Nor did it
sweep across the parent cloud. In fact, there is no known mechanism by which
low-mass star formation in one region can stimulate formation in another, distant
region. The accelerating star formation occurred globally. This process must have been
stimulated by contraction of the parent cloud. 

We mentioned previously that the present-day distribution of masses differs
somewhat from the canonical initial mass function. Returning to Figure~6, we
find that this distribution changed substantially as time progressed. Here,
the series of lower curves shows the cumulative mass function ${\cal N}_\ast$
at the indicated times. Thus, the bottom curve represents the number of stars
with mass less than any $M_\ast$ which are older than 5~Myr. Relatively few
stars have the requisite age. The figure shows, not surprisingly, that their
mass spectrum also does not resemble the standard initial mass function.

What determines the upper mass cutoff at each epoch? It is tempting to 
hypothesize that no relatively massive stars existed long ago because there had
not yet been sufficient time to form them. However, the figure itself suggests
an alternative explanation. The open circles show the masses for which the
pre-main-sequence lifetime equals the age in question. (Compare the dashed curve
of Figure 9.) At all times, the maximum mass is fairly close to this limit. In
other words, more massive stars could have indeed formed, but they already 
would have joined the main sequence. If such objects were (erroneously) 
assigned contraction ages, the latter would necessarily be less than the 
corresponding time.

\section{Physical Interpretation}
\subsection{Cloud Mass}
We have seen how the present number density of stars, $n_\ast (r)$, peaks at
a very high central value. Since the parent cloud was only recently dispersed,
its own density must have similarly peaked. The gas itself was not smoothly
distributed, but consisted of clumps, many of which already contained the stars
we see today. Averaging over such clumps, what might have been the radial
falloff in density and the total cloud mass, just prior to dispersal? 

A second clue comes from the observed proper motions of stars. Jones \& Walker 
(1988) found the vectors to be randomly oriented in the sky; the component in 
any direction has a mean magnitude of 
\hbox{$2.4\ {\rm km}\ {\rm s}^{-1}$}. The spread in proper motion magnitudes
is relatively small, about 0.2~km~s$^{-1}$.\footnote{There is a potential 
problem in calculating this dispersion for a sample of stars chosen by a 
membership criterion based on the proper motions themselves. However, Jones
\& Walker (1988) noted that the calculated dispersion is insensitive to the
precise membership criterion.} If the stars were indeed trapped in
the gas before dispersal, then we may interpret their speed as the 
one-dimensional velocity of the cloud gas. The observed randomness in 
orientation of the stellar velocities implies that the corresponding gas 
motion was similarly random. That is, the internal bulk motion of the cloud 
was turbulent. We equate the stellar proper velocity in any direction with 
$V_{\rm turb}$, the one-dimensional turbulent speed. 

It is conventional to model the momentum transfer associated with such
turbulent motion as arising from the pressure of an ideal gas. Then our cloud 
is effectively a self-gravitating, isothermal sphere. Because of the central 
peak, it is specifically the singular isothermal sphere, with a density given by
$$ \rho (r) \,=\, {V_{\rm turb}^2\over{2\,\pi\,G\,r^2}} \,\,, \eqno(2) $$
and a total mass of
$$ M_{\rm cloud} \,=\, {{2\,V_{\rm turb}^2\,R}\over G} \,\, \eqno(3) $$  
inside the radius $R$. For a cluster radius of 2.5~pc and the 
$V_{\rm turb}$-value taken from the stellar data, we find that 
\hbox{$M_{\rm cloud}\,=\,6700\ \Msun$}. Note that we have taken the cloud to be 
a full sphere, whose foreground half produced the visible stars. For 
comparison, the total mass of these stars is only $480\ \Msun$ for the sample 
of 244, and $580\ \Msun$ for the larger group of 705.\footnote{Hillenbrand
(1997) claims that her sample extends below a true turnover in the luminosity
function. Hence, our total mass estimate for the visible stars should be reasonably 
accurate.}

\subsection{Turbulent Dissipation}
Our estimate of the ONC cloud mass rests on the assumption that self-gravity
balanced turbulent pressure just prior to dispersal. How well is this assumption
supported by observation? It has long been known that the line widths of
optically thin tracers yield fluid velocities consistent with virial
values, over a large range of masses and sizes (Larson 1981; Myers \& Goodman
1988). This finding is generally interpreted to mean that clouds are indeed
supported by internal, turbulent motion.

Motivated by the observations of superthermal linewidths, a number of 
theorists have modeled, through direct numerical simulation, the dynamics of 
turbulence in a magnetized cloud gas. (See V${\acute{\rm a}}$zquez-Semadeni 
\etal 2000 for a review of such calculations.) Although the simulations 
adopted a variety of assumptions concerning the impressed turbulence, the 
results have been qualitatively consistent. In the absence of persistent 
driving, the turbulence decays rapidly, typically within a few crossing times.
This time is set by the average eddy speed and the size of the computational
box. Mac~Low (1999) has determined the energy dissipation rate from a suite of
MHD simulations. His essential conclusion is
$$ \dot\epsilon \,=\,  -\eta\,{V_{\rm turb}^3\over\lambda} \,\,.\eqno(4)$$
Here, $\dot\epsilon$ is the energy loss rate per unit mass of gas, 
$V_{\rm turb}$ the average (rms) eddy speed, and $\lambda$ the dominant 
wavelength of the impressed turbulence. The empirical, nondimensional 
coefficient $\eta$ was found to be about 0.4 in these simulations.  
 
These important experiments are actually finding {\it two} results;
both may not be applicable to real molecular clouds. The first, and more basic,
finding is that MHD turbulence is dissipative. Even incompressible modes 
(Alfv\'en waves) quickly transfer energy to compressive waves, which steepen 
and shock (Goldstein 1978; see also the discussion in Stone et al. 1998). This 
result is physically compelling, even if the actual dissipation in the 
simulations is numerical in origin.

The second result, more problematic astrophysically, is that the turbulence 
quickly decays. Here we must bear in mind that all the simulations are local.
Any global compression of the cloud due to self-gravity cannot be modeled.
But that very compression should supply energy and help sustain the turbulence. 
Indeed, if the mean eddy speed is to match the virial value at all times, as 
observations suggest, then that speed should {\it increase with time.} 
Furthermore, if the cloud contraction is mediated by turbulent dissipation, 
then this contraction naturally accelerates.

\subsection{Accelerating Contraction}
Let us illustrate these considerations through a simple, heuristic model, 
inspired by our reconstruction of the ONC cloud. The total energy, 
gravitational plus thermal, of the cloud in the recent past is that of the
bounded, singular isothermal sphere. For a cloud of mass $M_{\rm cloud}$ and 
radius $R$, this energy is given by
$$ E\,=\,-{1\over 4}\,{{G\,M_{\rm cloud}^2}\over R} \,\,.\eqno(5)$$

What was the cloud structure at previous times? The strong central peak in 
density was presumably attained just prior to the formation of the Trapezium, 
\ie, within the last $10^5$~yr. It is reasonable to assume that the cloud's 
density contrast monotonically increased with time to that point. If the 
turbulent velocity was also spatially homogeneous during earlier epochs, then 
the cloud evolved through a sequence of isothermal spheres, culminating in the
singular configuration.

The evolution is driven by self-gravity, mediated by the energy loss from 
turbulence. This bulk motion continually drives internal shocks, which 
themselves radiate energy at a {\it local} rate given by equation (4). If the
largest eddy has a size $\lambda$ comparable to the cloud diameter, then the 
mass-integrated heat equation is
$$ {{dH}\over{dt}} \,=\, 
-\eta\,{M_{\rm cloud}\,V_{\rm turb}^3\over{2\,R}}\,\,.\eqno(6)$$
The quantity $H$ is the cloud enthalpy, equal to 
\hbox{$E_{\rm tot} + P_\circ\,V$}. Here, $P_\circ$ is the bounding pressure, and
$V$ the cloud volume. As we show in the Appendix, it is this quantity whose
decrease yields the cloud's total, shock-generated luminosity. We regard $\eta$
as a free parameter, whose value we expect to be less than that found in the
current local simulations.

For the singular isothermal sphere, the enthalpy is
$$ H\,=\,-{1\over{12}}\,{{G\,M_{\rm cloud}^2}\over R} \,\,.\eqno(7)$$
Pending a more detailed calculation that tracks the cloud's changing internal
structure, we simply adopt this expression for all times, and substitute it into
equation (6). After utilizing equation (3) as well, we find an expression for
the decrease of $R$ with time:
$$ {{dR}\over{dt}} \,=\, -{{3\,\eta}\over{\sqrt{2}}} 
   \sqrt{{{G\,M_{\rm cloud}}\over R}} \,\,,\eqno(8)$$
which integrates to yield
$$ R \,=\, R_\circ 
\left(1\,-\, {{9\,\eta}\over{2\,\sqrt{2}}}\,{t\over t_\circ}\right)^{2/3}\,\,.
\eqno(9)$$
Here, $R_\circ$ is the present \hbox{($t\,=\,0$)} cloud radius. Equation (8) 
yields the cloud radius for each past epoch, \ie, for negative values of $t$.
The quantity $t_\circ$ is defined to be
$$
t_\circ \equiv \sqrt{{R_\circ^3\over{G\,M_{\rm cloud}}}} = 7.2\times 10^5\,{\rm yr} \eqno(10)$$

Equation (8) verifies that the cloud contracts in an accelerating fashion. 
Indeed, it undergoes a kind of diluted free fall. The time scale for this
process, some multiple of the free fall time $t_\circ$, is set by the (still
unknown) parameter $\eta$. As the cloud contracts, the speed $V_{\rm turb}$,
which is related to $R$ through equation (3), accelerates {\it upward}. This
conclusion is in striking contrast to the current numerical results 
(see V${\acute{\rm a}}$zquez-Sendani et al. 2000 for a review), which
show the turbulent speed to fall with time. The difference stems from our basic
assumption, motivated by the molecular line studies (\eg \ Myers \& Goodman 1988), 
that the cloud is nearly in virial equilibrium at all times. 

\section{Discussion}

Extending our earlier investigations of other systems, we have
documented empirically the acceleration of star formation in the ONC. 
We have shown that this process occurred at all stellar masses, so that there
is no present-day correlation between stellar age and mass. Furthermore, the 
acceleration was not confined to the cluster's central
region, but was global in character. We then surmised, on theoretical
grounds, that the parent cloud itself underwent an accelerating contraction,
again of a global nature, prior to its recent dispersal.

It is tempting, of course, to relate these two phenomena. Somehow, the
contraction of the ONC parent cloud must have stimulated the formation of
individual dense cores. This picture is self-consistent only if the dense
cores themselves quickly collapsed to stars. Such rapid evolution is
at odds with the traditional view that dense cores evolve over a time of
order $10^7$~yr through ambipolar diffusion (Shu et al. 1987). One intriguing
possibility is that ambipolar diffusion itself is enhanced by cloud turbulence 
(Nakamura \& Li 2005).

If star formation generally accelerates, then most molecular clouds should
exhibit relatively low rates of starbirth. Where are these quiescent objects?
Recall that most stars form within OB associations. The associations themselves
are produced by cloud clumps within giant complexes, like the parent entity
of the ONC. We suppose that these clumps evolve with time, gaining mass from 
their surroundings until they undergo accelerating contraction. Only the most 
massive clumps produce luminous clusters, as was found in the careful study of
the Rosette complex by Williams et al. (1995). But if the picture advocated 
here is correct, then sensitive infrared surveys should reveal lower levels of
star formation activity in the remainder of the clump population.

We have stressed that the Trapezium stars are truly anomalous in mass. More
specifically, Figure 4 shows that, for objects below about $3 \, M_{\odot}$, the ONC population is spatially homogeneous, but that its central region contains an unusual number of more massive
objects. This central concentration could not have been the result of 
dynamical relaxation in the brief interval since the dispersal of the
parent cloud (Bonnell \& Davies 1998). These facts argue for an alternative
formation mechanism for massive stars.

Both Bonnell \etal (1998) and Stahler et al (2000) have advocated coalescence models. 
The first authors pictured the merging units to be bare stars and their disks, while the second group invoked dense cores already containing young stars, which 
subsequently merge. Within the context of a globally contracting cloud, we
see how such cores, squeezed by the ambient pressure to sizes smaller than the
canonical 0.1 pc, would first coalesce in the central region. Theoretical
modeling of this phenomenon would be a welcome contribution. 

Finally, our study bears on the eventual fate of the ONC and similar groups. 
We have found that under 10~percent of the parent cloud mass consisted of 
stars, just prior to gas dispersal. Including obscured sources within
OMC-1 raises this figure. Even with this addition, the fraction is low enough that
the stars themselves cannot be bound solely by their mutual gravity. They will
disperse into space, forming a typically distended OB association. 

It has long been noted that many distant associations appear to contain tight 
stellar clusters at their centers (see, e.g., Garmany \& Stencel 1992). If
the ONC is typical in terms of star formation efficiency, then these groups
will similarly disperse. In contrast to this view, a number of theorists have 
proposed that bound clusters originate at the centers of OB associations
(Adams 2000; Kroupa et al. 2001). However, their calculations assume that a
large fraction (e.g., one third) of a clump's mass is converted to stars before
dispersal. In the traditional virial theorem analysis, the fraction necessary 
to produce a bound system is one half (Hills 1980). Our analysis of the ONC, 
the nearest rich cluster, indicates that these are overestimates. The origin of 
bound clusters must lie elsewhere. 

In a future study, we plan to follow in more detail the evolution of a 
cluster-forming cloud. We will track theoretically both the early growth of
such entities within a giant complex, as well as the structural changes that
accompany their final, accelerating contraction. We also hope to connect
quantitatively the rate of star formation at any epoch with the corresponding
cloud evolution. Such a calculation would be an important first step toward
understanding the larger issue of star formation efficiency on galactic 
scales.

\bigskip\noindent This project has been aided by illuminating discussions with
James Graham, Alessandro Navarrini, and Jon Swift. S. S. was partially 
supported by NSF Grant AST-9987266.
 
\appendix
\section{Heat Equation for an Isothermal Cloud}
Consider a mass element within a contracting, isothermal cloud of temperature
$T$. The Lagrangian changes in specific entropy, thermal energy, and mass 
density are related by

\begin{equation}
T\,\Delta s \,=\, \Delta\epsilon_{\rm therm} \,-\,
   {P\over\rho^2}\,\Delta\rho
\end{equation}

When we integrate over all such mass elements, the lefthand side of this 
equation becomes
\begin{equation}
T \Delta S=-L\Delta t
\end{equation}
where $S$ is now the total entropy, and where the luminosity $L$ is assumed to
arise from optically thin radiation. If we further invoke mass continuity, in 
the form
\begin{equation}
{{\Delta\rho}\over\rho} \,= -\left(\bnabla \bdot{\bf v }\right)  \Delta t
\end{equation}
then we find
\begin{equation}
-L\,\Delta t \,=\, \Delta E_{therm}+\Delta t \int\!{{dm}\over\rho}P\left(\bnabla\bdot {\bf v} \right)
\end{equation}
where $E_{\rm therm}$ is the total thermal energy. 

We recognize $dm/\rho$ as the volume element $d^3{\bf x}$. If we do an integration
by parts and evaluate the surface term, then the second righthand term in
equation (A4) becomes
\begin{equation} 
P_\circ\,\Delta V-\Delta t\int\!d^3{\bf x} \left({\bf v}\bdot \bnabla P\right)
\end{equation}
where $\Delta V$ is the total change in cloud volume. For the pressure gradient
$\bnabla P$, we employ Euler's equation
\begin{equation}
\bnabla P = -\rho\,\bnabla \Phi_{\rm grav}-\rho D{\bf v}/Dt
\end{equation}
where $\Phi_{\rm grav}$ is the gravitational potential. Further manipulation
transforms equation (A4) into
\begin{equation}
 -L \Delta t = \Delta E_{\rm therm} + P_\circ \Delta V +\Delta K \,+\, \Delta t\,
\int\!d^3{{\bf x}}\,\left({\bf v}\bdot\rho\,\bnabla \Phi_{\rm grav}\right)
\end{equation}
Here, $\Delta K$ is the change in the bulk kinetic energy:
\begin{equation}
K \equiv\, \int\!dm\,{v^2\over 2}
\end{equation}
Since we are not considering a rotating or pulsating cloud, we will henceforth
set $K$ equal to zero.

The remaining volume integral appearing in equation (A6) equals the change
in $E_{\rm grav}$, the gravitational potential energy. (See Tassoul
1979, p. 147, eq. (127), but note his sign error.) We thus find
\begin{equation}
-L\Delta t =\ \Delta E_{\rm therm} + \Delta E_{\rm grav} +
\Delta (P_\circ V)
\end{equation}

Dividing by $\Delta t$ yields the desired heat equation:
\begin{equation}
L = -{{dH}\over{dt}}
\end{equation}
where the generalized enthalpy
\hbox{$H\,\equiv\,E_{\rm therm}\,+\,E_{\rm grav}\,+\,P_\circ\,V$}. Note, 
finally, that for a monatomic gas, 
\hbox{$E_{\rm therm}\,=\,(3/2)\,M_{\rm cloud}\,a_T^2$}, where $a_T$ is the
isothermal sound speed. When evaluating $E_{\rm therm}$ for the singular
isothermal sphere, we have employed this relation, identifying $a_T$ with the
turbulent velocity $V_{\rm turb}$.

\end{document}